# Lossless broadband adiabatic polarizing beam splitting in a plasmonic system

Guang Yang, Alexander V. Sergienko, and Abdoulaye Ndao*

Department of Electrical and Computer Engineering & Photonics Center, Boston University, 8 Saint Mary's Street, Boston, Massachusetts 02215, USA

andao@bu.edu

## Abstract

The intriguing analogy between quantum physics and optics has inspired the design of unconventional integrated photonics devices. In this paper, we numerically demonstrate a broadband integrated polarization beam splitter (PBS) by implementing the stimulated Raman adiabatic passage (STIRAP) technique in a three-waveguide plasmonic system. Our proposed PBS exhibits >250 nm TM bandwidth with <-40 dB extinction and >150 nm TE bandwidth with <-20 dB extinction, covering the entire S-, C- and L-band and partly E-band. Moreover, near-lossless light transfer is achieved in our system despite the incorporation of a plasmonic hybrid waveguide because of the unique loss mitigating feature of the STIRAP scheme. Through this approach, various broadband integrated devices that were previously impossible can be realized, which will allow innovation in integrated optics.

## Introduction

Quantum physics processes can often be related to classical analogies in photonic systems [1] due to similar mathematical forms of the Schrödinger equation and the paraxial Helmholtz equation. The stimulated Raman adiabatic passage (STIRAP) technique [2, 3], originally developed for robust population transfer between three or more atomic or molecular energy levels, has found great applications in the manipulation and transfer of light in coupled waveguide systems. Since the initial proposal [4, 5] and experimental demonstration [6] of STIRAP in waveguides, variety of waveguide systems with versatile functionalities based on the STIRAP protocol have been reported, such as spectral filtering [7], beam splitting [8], and on-chip entanglement engineering [9]. Nevertheless, most current studies are limited to consideration of non-dissipative dielectric systems. The generalization of this approach to open systems creates new opportunities for utilizing plasmonics as well as the exceptional properties of non-Hermitian systems [10-12]. However, the adiabatic requirement of STIRAP, impacting the device length, hinders such endeavors and limits their practicality. The use of artificial gain is considered in recent reports to cancel the dissipation [13] or accelerate the transfer [14]. However, such conditions are impractical over the required distances that are typically centimeters long. In this letter we adopt an alternative approach inspired by a generalized atomic STIRAP framework, where the transfer can be lossless if the loss (decay) only exists in the intermediate state of the three-level system [15]. We exploit this unique feature and use a hybrid plasmonic waveguide (HPWG) [16] as the intermediate state in our waveguide system. This enables the implementation of a robust integrated polarization beam splitter (PBS) based on polarization selectivity of plasmonic system. The advantage of using STIRAP principles here is twofold: it circumvents plasmonic losses and provides excellent operation bandwidth.

With the rapid development of photonic integrated circuits, more and more degrees of freedom, such as wavelength and polarization, can be manipulated on the same chip. An integrated broadband PBS, preferably executed using high-index-contrast structures such as state-of-the-art lithium niobate on insulator (LNOI) platform [17-22], is a highly needed device. However, the operational bandwidth supported by conventional designs, i.e., directional couplers, is limited to tens of nanometers, owing to the wavelength sensitivity of the interference effect [23]. Extending the operational bandwidth requires more sophisticated engineering of anisotropic structures, such as hetero-anisotropic metamaterials [24] and cascaded interferometers with different birefringent paths [25]. While the performance of such devices is considered to be promising, the complexity of executing the sub-wavelength grating [24] and cascaded interferometers [25] is challenging and uncertainty in fabrication and device characterization increases. On the other hand, the STIRAP protocol has been utilized in simple dielectric-only waveguide systems and proved to offer large bandwidths [26-28]. However, these devices are based on low-index-contrast waveguide platforms that are incompatible with modern advanced integration technologies. Moreover, these dielectric-only designs require 3 to 5 cm long devices and, sometimes, at least a five-waveguide system [27], because their operation relies on the material birefringence only.

Here, we propose a novel broadband PBS in LNOI platform, which is based on a three-waveguide STIRAP scheme employing a plasmonic waveguide. To the best of our knowledge, this is the first STIRAP-based PBS reported in high-index-contrast integrated systems, featuring a short device length of 6 mm, which is 5 folds shorter than other STIRAP-PBS systems reported so far [26-28]. This approach opens fresh venues for on-chip polarization control and introduces new principles of plasmonic system design and manipulation.

## Design and operation principle

Fig. 1 presents a schematic of the proposed polarization beam splitter that illustrates its principle of operation. The device consists of three waveguides (WG1, WG2, WG3) made in X-cut thin-film lithium niobate ($h_{LN}$ = 375 nm) on top of a silicon dioxide layer. The outer two waveguides (WG1 and WG3) are fully etched ridge waveguides with the same width of $w_{LN}$ = 900 nm. The top cladding is air. Such parameters ensure single mode operation within the designed bandwidth (~1400–1700 nm). The HPWG in the middle (WG2) has a width of $w_{HPWG}$ = 300 nm and comprises three layers: an LN ridge waveguide, a gold topping layer ($h_{Au}$ = 100 nm), and a silicon dioxide interlayer (height g=80 nm) between gold and LN. The electric field in this HPWG is mainly confined in the low index silicon dioxide layer, in analogy to a slot waveguide [29-31], while the plasmonic nature of this structure imposes TM polarization on the resultant mode. For a TM mode excited in the left waveguide (WG1), the light transfer in the three-waveguide system is analogous to the STIRAP process in a three-level atomic system [fig. 1(b)], where the population initially in state |1⟩ (or WG1) is transferred to state |3⟩ (or WG3) via an intermediate state |2⟩ (or WG2). The intermediate state |2⟩ here is a decaying state (decay rate $\Gamma$), reflecting the lossy characteristic of WG2. In the waveguides system the detuning $\Delta$ reflects the propagation constant mismatch between WG2 and WG1/WG3. In contrast to the three-waveguide coupling for the TM case, a TE mode starting in WG1 undergoes direct two-waveguide coupling between WG1 and WG3, since WG2 does not support TE polarization. A PBS effect is achieved because such cross-coupling of the TE mode is weak given the large separation between WG1 and WG3. The focus of this study is the transfer of the TM mode, and all notations hereafter refer to TM unless otherwise specified.

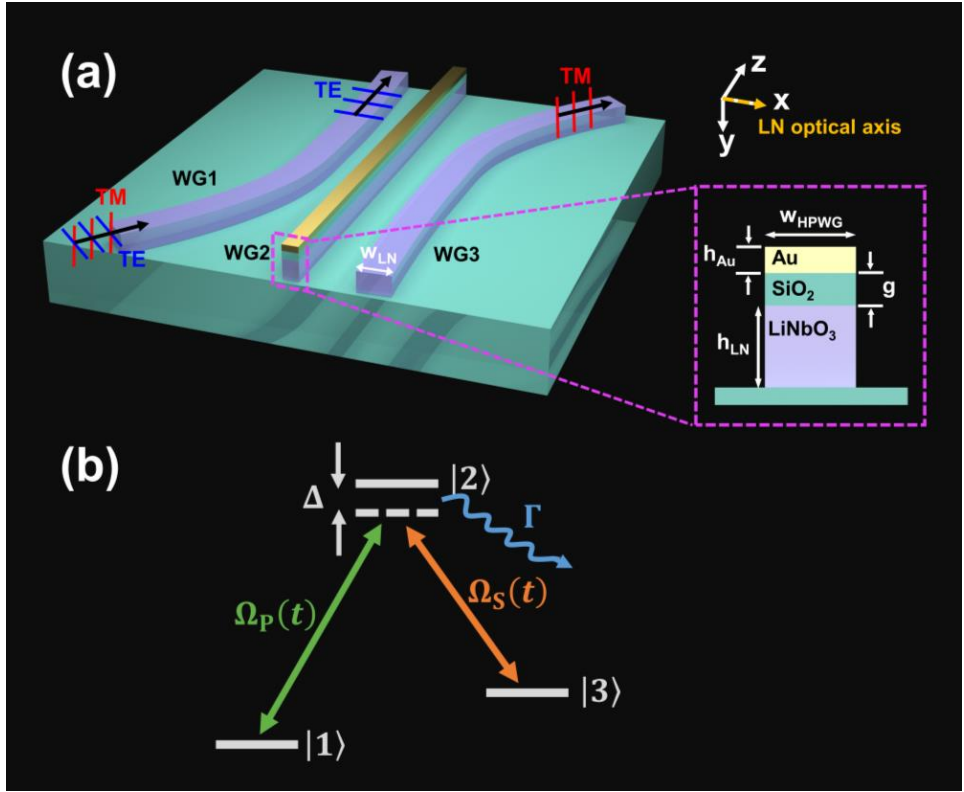

Fig. 1. (a) A schematic of the proposed device. Waveguide 1 (WG1) and Waveguide 3 (WG3) are identical in the cross section while bending in a centrally symmetric manner. The cross section of WG2 (the HPWG) is shown in the inset. (b) The three-level atomic system that is analogous to (a). The population transfer between states $|1\rangle$ and $|2\rangle$ is induced by the pump pulse $\Omega P$, and between states $|2\rangle$ and $|3\rangle$ by the Stokes pulse $\Omega S$. The pulses are represented as Rabi frequencies and are analogous to coupling strengths between waveguides.

We describe the state evolution in the waveguide system and highlight the analogy with a population transfer process based on STIRAP (substitution $z \to t$ illustrates the equivalency with an atomic system):

$$i\frac{\mathrm{d}}{\mathrm{d}z}\boldsymbol{A} = \mathbf{H}\boldsymbol{A}, \tag{1}$$

where $\boldsymbol{A} = [a_1(z)\ a_2(z)\ a_3(z)]^T$ denotes the field distribution in WG1, WG2 and WG3. The interaction Hamiltonian H in a perturbation form can be written according to the coupled mode theory (CMT) [32]:

$$\mathbf{H} \equiv \mathbf{H}^{(0)} + \mathbf{H}' = \begin{bmatrix} 0 & C_{12}(z) & 0 \\ C_{12}^*(z) & \Delta & C_{23}(z) \\ 0 & C_{23}^*(z) & 0 \end{bmatrix} + \begin{bmatrix} 0 & 0 & C_{13}(z) \\ 0 & i\Gamma & 0 \\ C_{13}^*(z) & 0 & 0 \end{bmatrix}, \tag{2}$$

The unperturbed Hamiltonian, H(0), represents a lossless system with detuning $\Delta$ in the intermediate level ($\Delta \equiv \Delta\beta = \beta_2 - \beta_1$ in the waveguide system), while the loss is accounted in the perturbation H′. $C_{ij}$ denotes the coupling coefficient between WG-i and -j, and $C_{ij} = C_{ji}^*$ since H(0) is lossless. The perturbation term H′ is responsible for two effects: the loss from WG2 and the coupling between WG1 and WG3. The latter is exclusive to the waveguide system, as the direct transition between $|1\rangle$ and $|3\rangle$ in the atomic system is forbidden. Following the perturbation approach, we analyze our system starting from the zeroth-order term, H(0), corresponding to a lossless STIRAP case. The full system is most conveniently understood when projected to the eigenstate basis of H(0), often referred to as the adiabatic basis:

$$\left|\Psi_+^{(0)}\right\rangle = \sin\phi \sin\theta \,|1\rangle + \cos\phi \,|2\rangle + \sin\phi \cos\theta \,|3\rangle$$

$$\left|\Psi_0^{(0)}\right\rangle = \cos\theta \,|1\rangle - \sin\theta \,|3\rangle \tag{3}$$

$$\left|\Psi_-^{(0)}\right\rangle = \cos\phi \sin\theta \,|1\rangle - \sin\phi \,|2\rangle + \cos\phi \cos\theta \,|3\rangle,$$

with eigenvalues $\lambda_0^{(0)} = 0$ and $\lambda_\pm^{(0)} = (\Delta \pm \sqrt{\Delta^2 + 4C_0^2})/2$, where $C_0(z) = \sqrt{|C_{12}(z)|^2 + |C_{23}(z)|^2}$. The parameters $\phi$ and $\theta$ are given by $\tan\theta(z) = C_{12}(z)/C_{23}(z)$ and $\tan 2\phi(z) = 2C_0(z)/\Delta$, respectively. These definitions lead to $\lambda_+^{(0)} \equiv C_0 \cot\phi$ and $\lambda_-^{(0)} \equiv -C_0 \tan\phi$.

Among the adiabatic states, of particular significance is $\left|\Psi_0^{(0)}\right\rangle$, also known as the dark state in STIRAP, since it does not involve the intermediate state $|2\rangle$. This highlights an important characteristic of ideal STIRAP: the population transfer takes place via $|2\rangle$ without populating $|2\rangle$ itself. In atomic STIRAP, the population transfer is enabled by the two pulses in the so-called counter-intuitive order, meaning the pulse connecting $|2\rangle$ and $|3\rangle$ precedes the pulse connecting $|1\rangle$ and $|2\rangle$. A similar configuration must be implemented in the waveguide system, such that for the initial state:

$$C_{12} \ll C_{23}; \quad \theta_i \approx 0; \quad \left|\Psi_0^{(0)}\right\rangle_i \approx |1\rangle, \tag{4}$$

and final state:

$$C_{12} \gg C_{23}; \quad \theta_f \approx \frac{\pi}{2}; \quad \left|\Psi_0^{(0)}\right\rangle_f \approx |3\rangle. \tag{5}$$

Thus, an initial population in $|1\rangle$ has perfect overlap with $\left|\Psi_0^{(0)}\right\rangle_i$ and can thereby evolve into $\left|\Psi_0^{(0)}\right\rangle_f$ (or $|3\rangle$), provided that it remains in $\left|\Psi_0^{(0)}\right\rangle$ without crosstalk to $\left|\Psi_+^{(0)}\right\rangle$ and $\left|\Psi_-^{(0)}\right\rangle$. However, as H(0) evolves, these instantaneous eigenstates become coupled, as can be seen from the following equation in the adiabatic basis:

$$i\frac{d}{dz}\begin{bmatrix} b_+ \\ b_0 \\ b_- \end{bmatrix} = \mathbf{H}_{adia}\begin{bmatrix} b_+ \\ b_0 \\ b_- \end{bmatrix}, \tag{6}$$

where the Hamiltonian in the adiabatic basis is given by

$$\mathbf{H}_{adia} = \begin{bmatrix} \lambda_+^{(0)} & i\dot\theta \sin\phi & i\dot\phi \\ -i\dot\theta \sin\phi & \lambda_0^{(0)} & -i\dot\theta \cos\phi \\ -i\dot\phi & i\dot\theta \cos\phi & \lambda_-^{(0)} \end{bmatrix} + \begin{bmatrix} -i\Gamma\cos^2\phi & 0 & i\Gamma\sin 2\phi/2 \\ 0 & 0 & 0 \\ i\Gamma\sin 2\phi/2 & 0 & -i\Gamma\sin^2\phi \end{bmatrix}. \tag{7}$$

In Eq. (7), an over dot denotes z-derivative. Here we have again used the perturbation approach and dropped the $C_{13}$ terms for simplicity. It is clear that $\mathbf{H}_{adia}$ is not exactly diagonal despite the instantaneous eigenstate basis. However, if the system evolves slowly, the derivatives in the off-diagonal terms will vanish, and the modes are nearly decoupled. This particularly leads to isolation of the $\left|\Psi_0^{(0)}\right\rangle$ state, since its related elements in the perturbation matrix are zero. Therefore, near-lossless evolution within $\left|\Psi_0^{(0)}\right\rangle$ can be obtained, although the exact lossless limit requires infinite interaction length. Note that the $\left|\Psi_n^{(0)}\right\rangle$ states are the zeroth-order terms of the exact eigenstates of H. The latter can be expressed by the perturbation theory:

$$|\Psi_n\rangle = \left|\Psi_n^{(0)}\right\rangle + \sum_{m \neq n} \frac{\left\langle \Psi_m^{(0)} | \mathbf{H}' | \Psi_n^{(0)} \right\rangle}{\left(\lambda_n^{(0)} - \lambda_m^{(0)}\right)} \left|\Psi_m^{(0)}\right\rangle, \qquad m, n \in \{+, 0, -\}, \tag{8}$$

to the first-order approximation. These exact eigenstates will be visualized in Fig. 2 (b)-(d).

## Device simulation

To implement the STIRAP-like light transfer in a waveguide system, we consider the waveguide layout depicted in Fig. 2(a). The shapes of WG1 and WG3 are defined by $x_{\text{WG1}}(z) = -a_0|z - z_{0\text{WG1}}|^3 + d_{\min} + (w_{\text{LN}} + w_{\text{HPWG}})/2$ and $x_{\text{WG3}}(z) = a_0|z - z_{0\text{WG3}}|^3 + d_{\min} + (w_{\text{LN}} + w_{\text{HPWG}})/2$, respectively, where xWG1(z) and xWG3(z) measure the z-dependence of the center positions of WG1 and WG3. dmin = 700 nm is the minimum edge-to-edge distance, occurring at $z_{0\text{WG1}}$=4.5 mm for WG1 and $z_{0\text{WG3}}$=1.5 mm for WG3. The full device length is 6 mm. The coefficient affecting the degree of bending is $a_0$=16.33 when all variables are converted to the unit of meters. These shapes are designed to obtain super-Gaussian-shaped coupling coefficient profiles [see Fig. 3(a)]. Despite the noticeable bending in Fig. 2(a), the local directional angle between the waveguide axial direction and the z-axis is very small (dx/dz=0.001 at maximum). Therefore, the refractive index change resulting from the propagation direction in X-cut LN is negligible. This also makes it convenient to use the local normal mode representation to illustrate the TM mode evolution in the waveguide system, as presented in Fig. 2 (b)-(d). We used a 2D finite difference mode solver (Lumerical MODE) to obtain the supermodes. The simulation wavelength is 1550 nm. The refractive indices of LN are 2.1379 (extraordinary) and 2.2112 (ordinary) [33], and the indices of gold and silicon dioxide are 0.5301+10.81i [34] and 1.444 [35], respectively. The supermodes are correlated to the adiabatic states in Eq. (8) [or Eq. (3) with zeroth-order approximation], since both representations depict the same eigenstates of H. The evolution of these adiabatic modes during the field propagation can be clearly seen from (b) to (d) in Fig. 2. Particularly, the $|\Psi_0\rangle$ state, which evolves from $|1\rangle$ into $|3\rangle$, is observed to be a dark state not involving WG2. This explains the near-lossless light transfer mechanism, even if a lossy HWPG is included in the system.

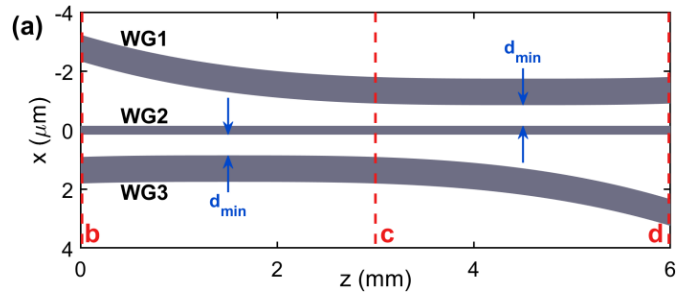

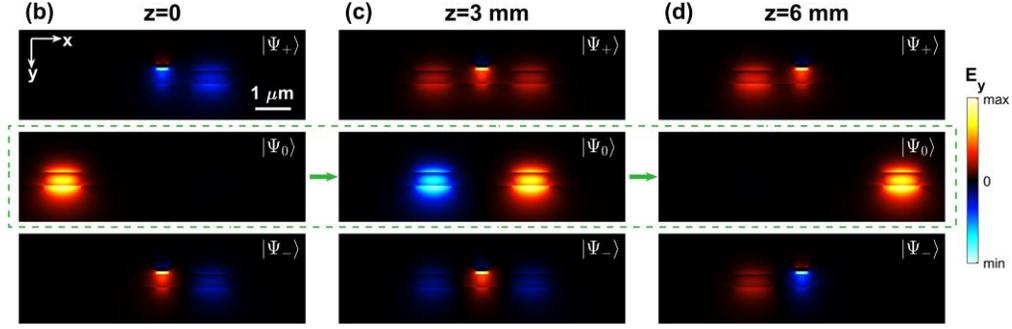

Fig. 2. (a) The layout of the three-waveguide system. (b)-(d) The electric field of the TM supermodes (instantaneous eigenstates) at the cross section of (b) z=0, (c) z=3 mm, (d) z=6 mm. The corresponding cross sections are designated in (a). For the STIRAP-like transfer of the TM mode, the system evolves mainly in the $|\Psi_0^{(0)}\rangle$ state in the dashed box. For TE supermodes, see Appendix A.

To examine the performance of the proposed PBS, we conducted CMT calculations and 3D simulations using the eigenmode expansion (EME) method (Lumerical EME). For CMT, we extract the coupling coefficients between two waveguides following the derivations in [36]. The resultant coupling coefficient profiles along z are plotted in Fig. 3(a). Empirically, the coupling coefficient decays exponentially with the waveguide separation. Thus C12(z) and C23(z) take the form of truncated third-order super-Gaussian functions resulting from the "reflected" cubic function profiles of WG1 and WG3. Such profiles are designed to optimize the overlap of the coupling strengths for STIRAP transfer. To provide guidance for the optimization, we plot two parameters in Fig. 2(b): the angle $\theta$ in Eq. (3), and an adiabatic condition parameter defined as $\sqrt{|C_{12}(z)|^2 + |C_{23}(z)|^2}/\dot{\theta}$. This parameter compares the diagonal elements ($\sim\sqrt{|C_{12}(z)|^2 + |C_{23}(z)|^2}$) in Eq. (7) to the off-diagonal elements ($\sim\dot{\theta}$) and estimates the degree of adiabaticity. It is evident in Fig. 3(b) that $\theta$ evolves smoothly from initial value 0 to final value $\pi/2$, corresponding to $|1\rangle \rightarrow |3\rangle$, and the adiabatic condition number >100 holds for the entire propagation, indicating the transfer here is close to the adiabatic limit. We calculate the fractional power in the waveguides following Eq. (1) and Eq. (2). The CMT result of TM input from WG1 is shown in Fig. 3(c). High transfer efficiency of 91.2% from WG1 to WG3 is achieved, and the residual power in WG1 is negligible (~10-6). The corresponding EME simulation is presented in Fig. 3(d). The simulated efficiency of 93.2%, extracted from the scattering matrix, is in good agreement with CMT. For the TE mode, the two-waveguide coupling can be expressed by

$$i\frac{d}{dz}\begin{bmatrix}a_{1TE}\\a_{3TE}\end{bmatrix} = \begin{bmatrix}0 & C_{13TE}(z)\\C_{13TE}^*(z) & 0\end{bmatrix}\begin{bmatrix}a_{1TE}\\a_{3TE}\end{bmatrix}. \tag{9}$$

The calculated fractional power is plotted in Fig. 3(e). Due to the small direct coupling coefficient between WG1 and WG3 [see Fig. 3(a)], the cross coupling from WG1 to WG3 is <1% (-20 dB). This is also validated by the EME simulation in Fig. 3(f), which shows near unity transmission through WG1. Complete discussion on the TE transfer and device length scaling is provided in Appendix A and B. The transfers here are reciprocal, thus beam combining of TE and TM can also be obtained.

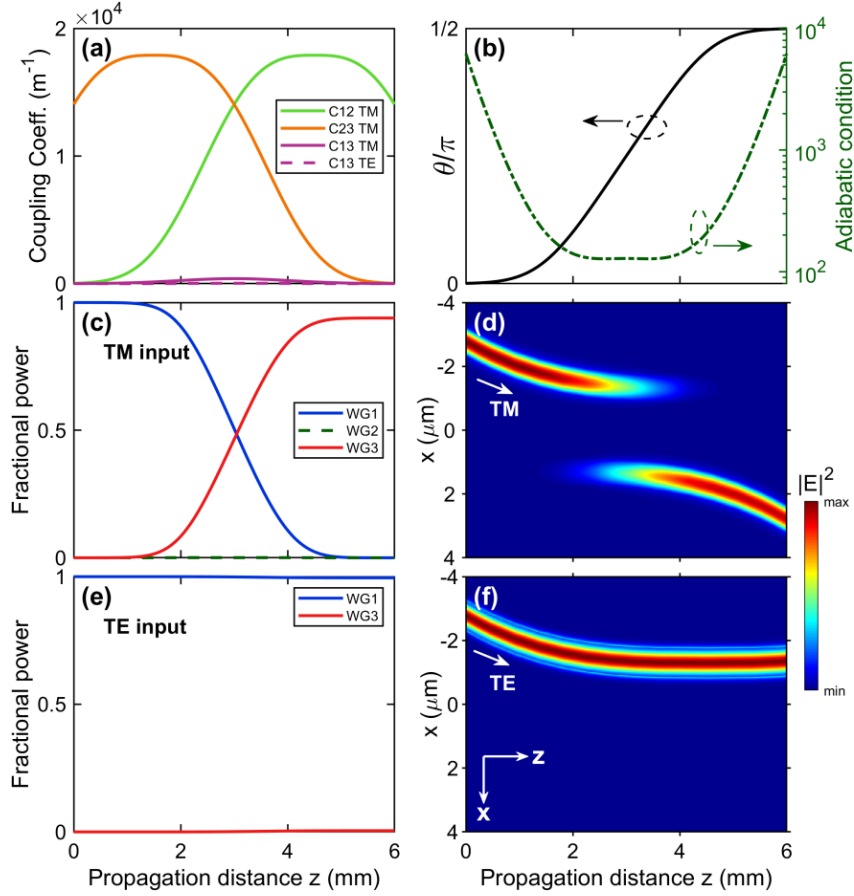

Fig. 3. (a) The coupling coefficient profile. (b) The evolution of the parameter θ and the adiabatic condition defined in the text. (c), (d) Propagation of a TM mode injected in WG1 by (c) CMT calculation and (d) EME simulation. (e), (f) Propagation of a TE mode injected in WG1 by (e) CMT calculation and (f) EME simulation. (d) and (f) represent the intensity.

The STIRAP process is well-known for its robustness against parameter deviations. This is because the transfer relies on the relative strengths of the coupling coefficients (or the pulses in atomic STIRAP), rather than their absolute values. As a result, our PBS exhibits a large bandwidth and good tolerance to fabrication imperfections. Fig. 4 (a) and (b) present the simulated bandwidth of the PBS with target parameters. The extinction of the TM mode transfer is <-40 dB (at the output of WG1, or bar port) over a bandwidth of ~250 nm (1450–1700 nm). The insertion loss is 0.3 dB at the design wavelength (1550 nm), and stays less than 1 dB for approximately the same bandwidth interval (1425–1675 nm). For the TE mode, the cross coupling to WG3 (cross port) becomes more significant at larger wavelengths. Nonetheless, the extinction is still below -20 dB for wavelengths shorter than 1550 nm. A significant source of imperfection in current LNOI fabrication technologies is slanted sidewalls. For the design, we have assumed 90° sidewalls for simplicity. Here we also examine the case with 80° sidewalls ($\Delta\theta = 10°$) in Fig. 4 (c) and (d). The extinction of TM (<-30 dB) is affected very slightly, whereas the bandwidth, in this case, can extend to even shorter wavelengths. Another delicate parameter is the height of the silicon dioxide interlayer in the HPWG. We show in Fig. 4 (e) and (f) that $\Delta g=10$ nm would not affect the extinction at 1550 nm. In both cases, the effect of fabrication imperfections on the TE performance is minimal (for complete tolerance analysis, see Appendix C).

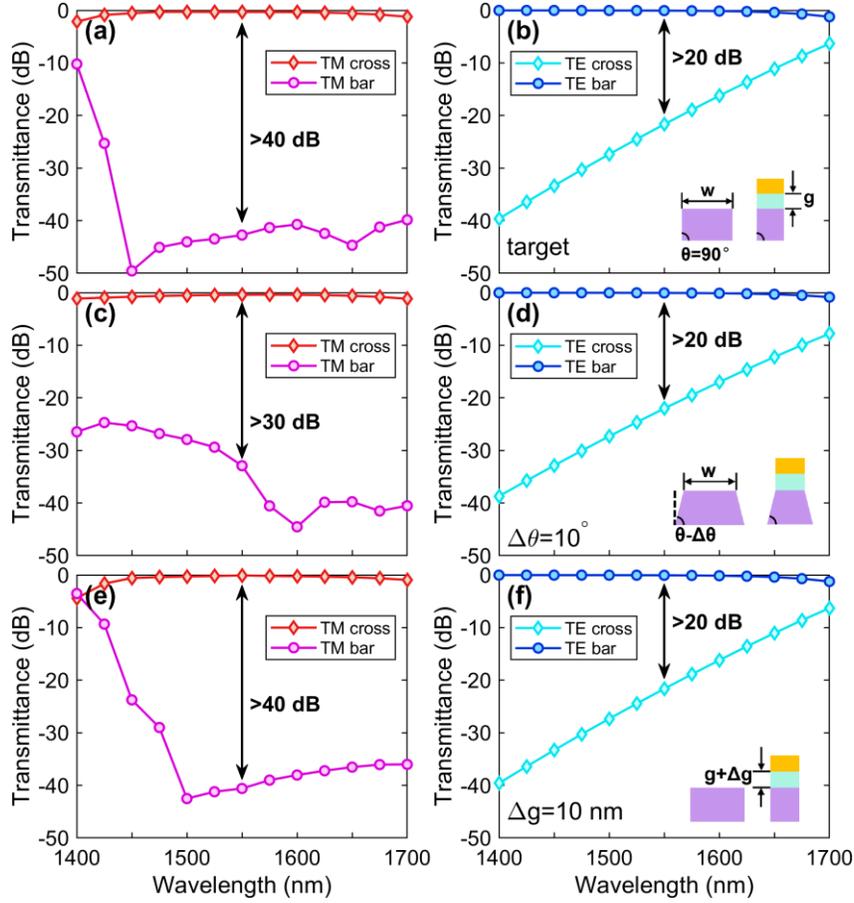

Fig. 4. Simulated transmittance spectra of the PBS with (a), (b) target parameters; (c), (d) slanted sidewalls (Δθ=10°); (e), (f) deviation of the thickness of the SiO2 interlayer (Δg=10 nm). (a), (c) and (e): TM excitation in WG1. (b), (d) and (f): TE excitation in WG1. The cross port refers to WG3; the bar port refers to WG1.

## Conclusion

In conclusion, we have investigated the waveguide analogy of a STIRAP system involving a decaying intermediate state and proposed a novel design of integrated PBS. The STIRAP scheme enables broadband light transfer in a three-waveguide system via a plasmonic waveguide. The inherent losses from the plasmonic waveguide are circumvented thus achieving a low-loss broadband plasmonic PBS. We theoretically and numerically demonstrated a robust transfer (<-40 dB extinction) of the TM mode over ~250 nm bandwidth, covering the entire S-, C- and L-band and partly E-band, with TE extinction <-20 dB for wavelengths <1550 nm. Moreover, the insertion loss is <1 dB almost throughout the bandwidth is achieved despite the presence of a plasmonic structure. Our proposed PBS is compatible with the state-of-the-art LNOI platform and facilitates the development of integrated photonic circuits.

## Acknowledgement

This research was supported by the National Science Foundation EFRI-ACQUIRE Grant No. ECCS-1640968, AFOSR Grant No. FA9550-18-1-0056, and DOD/ARL Grant No. W911NF-20-2-0127 and the financial support of Boston University start-up funding.

## Appendix A. TE supermodes and TE coupling

As stated in the paper, the TE coupling is outside the STIRAP framework. Since the hybrid plasmonic waveguide (not visible in Fig. 5) does not support a TE mode, the TE coupling is a two-level problem, and only two TE supermodes are associated with the system. Fig. 5 plots the TE supermodes at a sample plane of z=3 mm. The TE coupling follows the manner of a two-waveguide directional coupler and can also be seen as the multimode interference between the symmetric and anti-symmetric supermodes. This coupling has more wavelength dependence (as all interference effects) than the STIRAP transfer of the TM mode, but overall, the coupling is weak because of the large gap between WG1 and WG3.

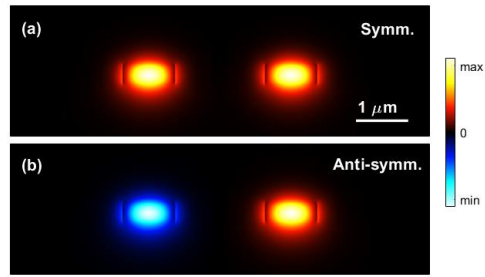

Fig. 5. The electric field of the two TE supermodes at z=3 mm. Note that these modes do not involve plasmonics. The simulation is conducted at the wavelength of 1550 nm.

## Appendix B. Device length scaling

Here we discuss considerations regarding scaling device length. Fig. 6 presents the dependence of the transfer efficiencies on the device length. In this analysis, only the device length is scaled, and the relative shapes of the waveguides are kept the same. From Fig. 6, it can be seen that the TM coupling efficiency grows with the device length, although the increment slows down in the regime where the adiabatic condition is well met (for sufficiently long devices). As pointed out in the paper, complete lossless transfer requires infinite device length. If higher efficiency is desired, the device length can be scaled to approach the ideal limit. In the paper, we have chosen a length of 6 mm to balance device footprint and efficiency. It is also worth pointing out that our device does not exhibit strong device length dependence, which highlights its adiabatic evolution nature [37], in contrast to directional-coupler-based PBS devices that rely on interference effects, such as the one in [38], which exhibits a seemingly similar structure but is based on different principle of operation. We will discuss further the different physics behind under Appendix D. It can also be manifested from this analysis that STIRAP is an intrinsically robust protocol. Once the adiabatic condition is met, it is insensitive to various parameter deviations, and does not require particular engineering of dispersion [39] to achieve broadband performance.

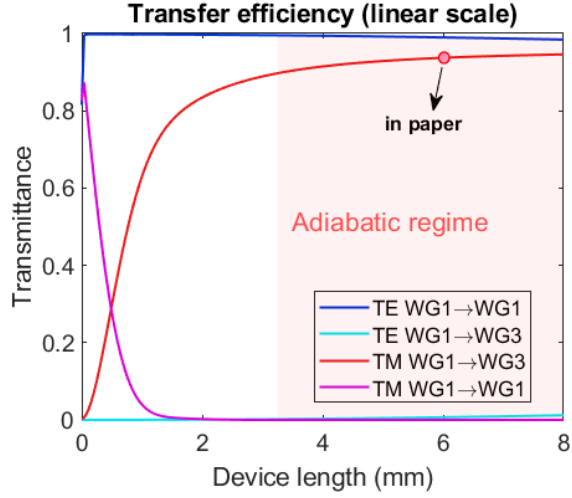

Fig. 6. Simulated transfer efficiencies as a function of the device length. Note that the analysis here is scaling the device, rather than the propagation in a fixed-length device. The simulation is conducted at the wavelength of 1550 nm.

## Appendix C. Complete analysis of error tolerance

In addition to the representative cases of fabrication error analysis in Fig. 4 of the paper. Figs. 7, 8, and 9 present a complete analysis of all tolerances.

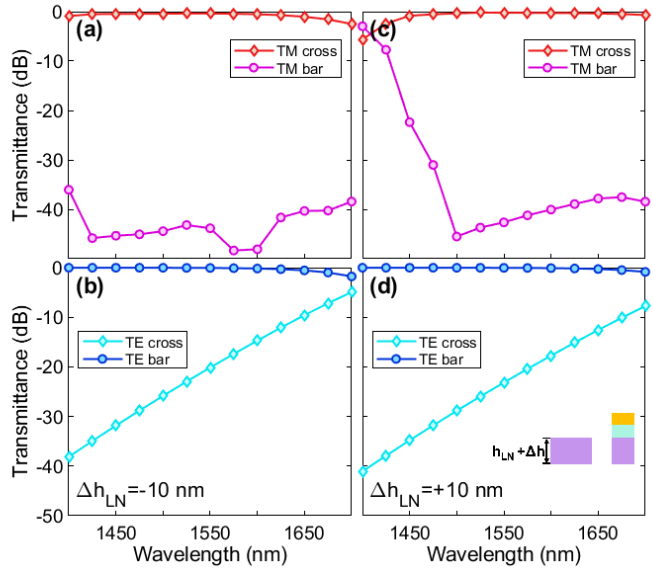

Fig. 7. Simulated transmittance spectra of the PBS with deviation of LiNbO3 layer thickness. (a), (b) $\Delta h_{LN} = -10$ nm; (c), (d) $\Delta h_{LN} = +10$ nm. The cross port refers to WG3; the bar port refers to WG1.

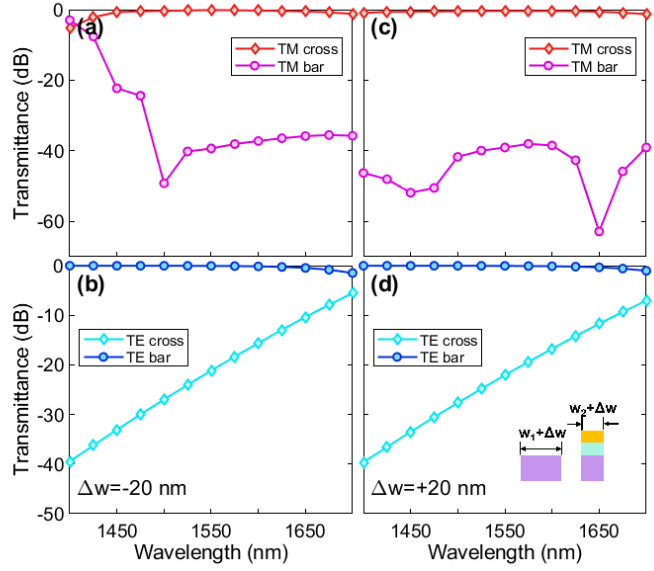

Fig. 8. Simulated transmittance spectra of the PBS with deviation of waveguide width. (a), (b) Δw= –20 nm; (c), (d) Δw= +20 nm. Note the different vertical axis scale of TM spectra [(a) and (c)].

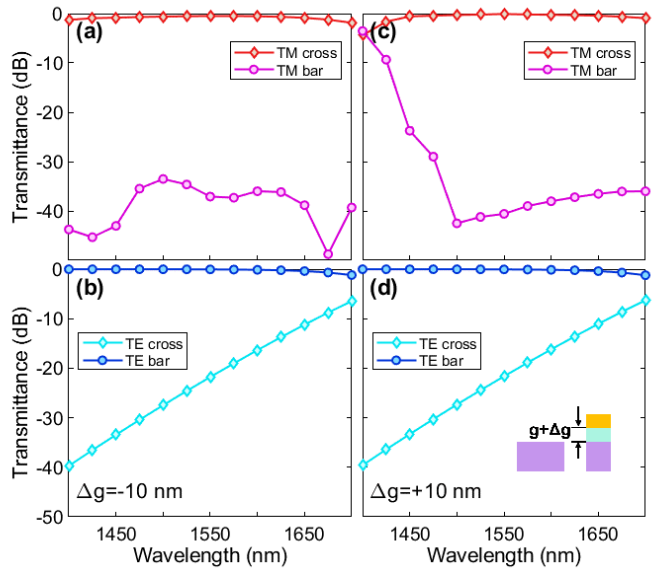

Fig. 9. Simulated transmittance spectra of the PBS with deviation of the gap (SiO2 interlayer). (a), (b) Δg= –10 nm; (c), (d) Δg= +10 nm.

**Appendix D. Unique physical principle of the STIRAP device.**

In Appendix B we have addressed the uniqueness of the STIRAP PBS from the perspective of device scaling. Below we present a detailed comparison between a directional-coupler-based PBS, such as the one proposed in [38], and our device, despite their seemingly similar structures.

Table 1. Principle comparison between a directional-coupler-based PBS and a STIRAP PBS.

|  | Directional-coupler-based PBS ([38]) | STIRAP-based PBS (this work) |
|---|---|---|
| Evolving coupler profile | No | Yes |
| Operation principle | Directional coupler | STIRAP |
| Interference effect | Yes | No |
| Mode evolution effect | No | Yes |
| Excited supermodes | All three supermodes | Only the $|\Psi_0\rangle$ mode |
| Dissipation due to excitation of plasmonic mode | Yes | No in the ideal limit; Yes with finite device length |
| Device length requirement | Phase matched coupling length; short but specific | Adiabatic condition; long but insensitive |

Particularly worth noting is the fact that in [38], the plasmonic mode is excited, as can be manifested by the appreciable plasmonic field in Fig. 5(a) therein. This is in direct contrast to our work, where the plasmonic waveguide remains dark throughout propagation [see Fig. 4(d)]. This illustrates that the $|\Psi_0\rangle$ mode we utilize here is a dark mode. This dark-plasmonic mode depicts the underlying reason for the loss-immune performance.